\documentclass[prb,preprint]{revtex4-1} 

\usepackage{amsmath}
\usepackage{amsfonts} 
\usepackage{graphicx} 
\usepackage{multirow} 

\usepackage{tabu}
\graphicspath{ {images/} }

\begin{document}
\title{On the determination of accelerometer positions within host devices}
\author{Christopher Isaac Larnder}\email{chrisisaac.larnder@johnabbott.qc.ca}
\author{Brian Larade}
\affiliation{Department of Physics, John Abbott College, St-Anne-de-Bellevue QC, Canada H9X 3X8}

\begin{abstract}

A review of recent accelerometry experiments points to the need for a careful consideration of the question of where, exactly, the accelerometer sensor itself is located within the device that hosts the services required for its operation. We propose a simple measure for characterising the regime of experimental conditions under which these implied positional errors become significant. We then present a method for reliably determining the accelerometer position and apply it to a wide variety of host devices ranging from data loggers to smartphones and tablets. Although the obvious value of the method lies in the improved accuracy of the ensuing accelerometry, the preliminary process of discovering the accelerometer position becomes a valuable and complete pedagogical experience in itself, providing insight into both the principles of radial acceleration and the internal structure of current-generation digital devices.

\end{abstract}

\maketitle

\section{Introduction}
\label{Introduction}
Low-cost motion sensors such as accelerometers are ubiquitous in today's digital landscape of handheld devices and "smart" technologies, and embedded in a wide variety of everyday objects\cite{sensors}  \cite{actuators} \cite{ieee}. It is natural that they have been increasingly adopted as tools for investigating motion\cite{hinrichsen} and, in particular, for laboratory experiments supporting physics education\cite{kuhn}\cite{ochoa}.

\subsection{Host Devices}
\label{Host Devices}
Although micro-eletromechanical systems (MEMS) such as current-generation accelerometers are celebrated for their small size\cite{mems}, little mention is made of the fact that these chips do not function without a supporting hosting environment that provides, minimally, a source of energy; a means of storing or wirelessly transmitting data; control logic; and at least some basic user-interface elements. Often enough, for example in today's smart phones, these components come bundled together with many other systems and services. All these components, together with their shared protective housing, constitute the much larger host device with which accelerometry is actually performed.

A question naturally arises, then, regarding the need for not only the position of the host device with respect to the physical system being studied, but also for the position of the accelerometer sensor itself within the volume of space occupied by the host device. Although the natural expectation is that this latter position would correspond to the exact center of the host device, other design considerations appear to take precedence: the sensor is rarely found at the geometrical center of the device. Prior to developing procedures for determining this intra-device position, however, should be a consideration of the experimental conditions under which we expect this level of detail to become significant.

\subsection{Linear vs. Rotational Motion}
\label{Linear vs. Rotational Motion}
Regardless of the device used, motion investigations involving accelerometers can be classified into two broad categories: linear motions and rotational motions. For linear motions, the acceleration of all points within the rigid enclosure of the host device are identical, and thus the actual position of the sensor within that volume is irrelevant. During rotational motions, however, every point within the device experiences a distinct vector of acceleration \cite{goldstein}\cite{footnote_rot}. In these cases, the exact position within the device may be important, ignorance of which would lead to geometrical uncertainties on the order of the dimension of the device, which can range from a few centimeters for small data-logger type devices, on the order of 10 cm for current-generation smart phones, and to another factor of at least 3 for consumer tablets.

The experimental significance of this device-level uncertainty depends, in turn, on the characteristic length scales associated with the motion being studied. In the case of inertial guidance systems \cite{inertial_guidance} \cite{footnote_gyro}, for example, the relevant motion parameter would be the radius of curvature of the airplane's motion path whose change in value due to taking into account the actual position of the sensor within the enclosure of the host device would be undetectably small. 

Generally speaking, the true intra-device position can be ignored when

\begin{equation}
\label{eq1}
\epsilon  \equiv \frac{L\textsubscript{device}}{R\textsubscript{motion}} << 1
\end{equation}

in which we define a significance parameter $\epsilon$ as the ratio of $L\textsubscript{device}$, the length scale associated with the size of the host device, to $R\textsubscript{motion}$ , a (radial) length scale characteristic of the motion being examined. The position-independent acceleration of linear motion can be found by noting that $\epsilon\rightarrow$0 in the large $R\textsubscript{motion}$ limit.

\subsection{Rotational accelerometry in laboratory settings}
\label{Rotational accelerometry in laboratory settings}
Although inequality \eqref{eq1} holds well for airplane motion, this is not always the case for motions that are confined to a laboratory. Given that the number of papers pertaining to the use of accelerometers in physics education has increased significantly in recent years; that the length scales ($R\textsubscript{motion}$)  in this domain are typically on the order of the size of a table-top, at most; that the most common device employed is the smartphone, whose dimensions ($L\textsubscript{device}$)  are  rather large compared to other available host devices\cite{footnote_dims} ; and that the market trend in recent years is towards still larger-sized phones, the question of intra-device sensor position is becoming increasingly relevant. A concerted discussion, however, has yet to emerge from recent literature.

One paper, for example, reporting results of a smartphone rotating on a merry-go-round\cite{monteiro1} with radius values as small as 40 cm, refers only to the .001 m/s\textsuperscript{2} accuracy of the sensor itself without considering the uncertainties due to the unknown intra-device position.  A study involving a smartphone on a rotating platform\cite{brasileiro} at a nominal radius of 82 mm claimed that the sensor was located exactly at the center of the phone. A similar study using a Nintendo Wiimote\cite{ochoa} assumed  the location of the accelerometer to be, within an uncertainty of 5 mm, under the centrally-located ``A'' button of the device, when the actual location is closer to 14 mm off-axis from this reference position\cite{footnote_wiimote}. An experiment in which a cellphone is attached to a vertically-rotating bicycle wheel\cite{monteiro2} states, without further comment, that their radial distance of 30 cm  was measured using the center-of-mass of the cellphone as the reference position. One paper examining a spinning roof slat\cite{vogt} does acknowledge, in a footnote, the role of intra-device position on accuracy, but does not take steps to account for it. 

The authors of a 2014 rotating-disc study\cite{hochberg} are the first to propose a method for finding the sensor position, the reasons for which are clear when we note that their range of radius values (as small as 1.4 cm) were on the order of and sometimes smaller than the size of the host device itself, producing greater-than-unity values for parameter $\epsilon$ of Eq. \eqref{eq1}. Their method, however, is a trial-and-error approach in which the device is repeatedly moved about near the center of rotation until all (horizontal) acceleration components read a value that is satisfactorily close to zero; and they do not, in any case, discuss the termination condition used for this iterative process nor report the final position obtained. 

In the last section of a 2015 survey article on smartphone experiments\cite{vieyra} appears the first formal albeit mostly geometrical method for finding the sensor position involving the intersection of two arcs. In the only other paper\cite{mau} that has, to the author's knowledge, been published on the subject, the arc-intersection method is combined with a method of intersecting radial lines. 
 
No other authors appear to have refined, adopted nor even become cognizant of any of the three methods just mentioned, which is understandable given that these methods were, for the most part, a secondary detail to the papers' principal topics. Indeed, the 2015 paper does not reference the 2014 one, nor does the last paper, appearing in 2016, refer to either of the other two. The current level of awareness is exemplified by an otherwise very methodical 2017 door-slamming study\cite{klein} published in the present journal, which does not even refer to the sensor-position problem in their use of an iPod Touch, estimating a radial uncertainty of 1 mm while the true uncertainty due to ignorance of the intra-device sensor position is likely an order of magnitude larger\cite{footnote_klein}.

The present paper adopts an algebraic method in which the radial acceleration vector for each individual device position produces an independent estimate of the intra-device sensor position. By accumulating estimates from multiple trials, we obtain both increased statistical reliability as well as an estimate for the uncertainty of the averaged position.

\subsection{Theory}
\label{Theory}
We idealize accelerometer sensors as point objects that can be associated with a singular position $\vec{R}$, and assume that the readings for each axis correspond to the true acceleration values at exactly that position. First-order corrections to this model depend on the specific engineering strategies employed by current-generation microelectromechanical systems, to which category today's accelerometers belong. Regardless of the details, these corrections would involve some type of averaging process over surface areas with linear dimensions that range anywhere from 1 $\mu$m to at most 1 mm\cite{sensordims}.

The dimensions of the host device, on the other hand, are necessarily and considerably larger. We will assume that they dominate experimental errors related to inaccurate knowledge of sensor position, i.e.

\begin{equation}
\label{eq2}
L\textsubscript{sensor} << L\textsubscript{device}, 
\end{equation}

where L\textsubscript{sensor} and L\textsubscript{device} are the typical length scales associated with the sensor and with the host device, respectively.

In what follows, we will restrict ourselves to experimental conditions requiring the analysis of accelerations in only two dimensions. 

\begin{figure} [ht]
\includegraphics[width=7 cm]{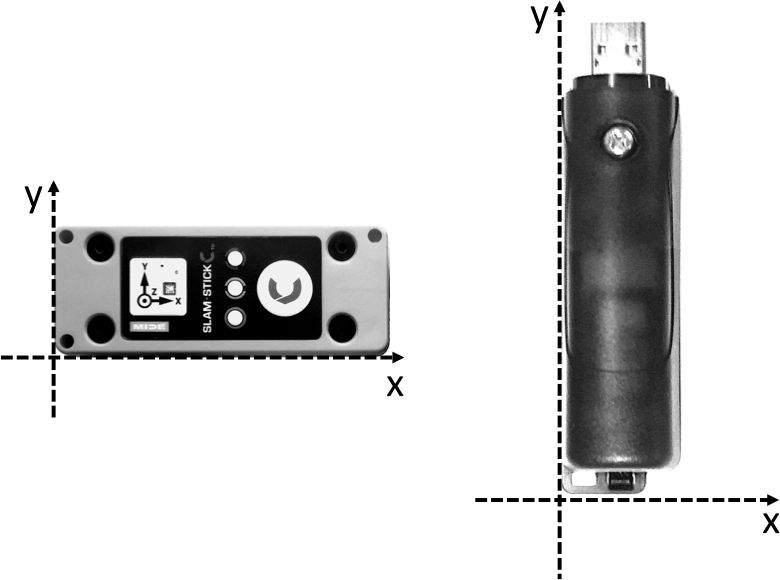}
\centering
\caption{Illustration of local device coordinates for two host devices. ( left: MIDE SlamStick C; right: GCDC X2-2 ) The choice of origin is derived from the manufacturer's choice of x- and y-coordinates reported by the accelerometer.}
\label{fig:LocalCoordinates}
\end{figure}

In order to associate a local coordinate system with the host device we note, with little loss in generality, that all the devices we investigated had a clear rectangular shape, establishing the orientation of the major axes; that the orientation of the accelerometer aligns itself with these axes; that the manufacturer already establishes, for each axis, which direction corresponds to a positive reading; and that there is an obvious "up-facing" surface allowing us to unambiguously refer to the "bottom-left" corner of any given host device, a reference point from which all vector positions within the device will have positive components\footnote{To make this rule consistent with host devices running iOS requires a preprocessing step in which all acceleration values are multiplied by -1}.  We therefore propose a convention in which this corner is used as the origin for all local device coordinates, as illustrated in Fig.~\ref{fig:LocalCoordinates} Accordingly, we denote $\vec{r}$ as the 2-D idealized-point location of the accelerometer sensor in this local device coordinate system, a constant implicitly set by the manufacturer, and whose experimental determination is the principal subject of this paper. 

In order for the accelerometer to provide information about a physical system, it will have to be rigidly attached to an object. We will restrict our discussion to such an object undergoing circular motion. 

For a system undergoing uniform circular motion at angular velocity $\omega$, elementary vector calculus demonstrates that the acceleration $\vec{a}$ at position $\vec{R}$ is given by

\begin{equation}
\label{eq3}
\vec{a} = -\omega^2 \vec{R}, 
\end{equation}

where $\vec{R}$  and $\vec{a}$ are vectors in a 2-dimensional planar coordinate system whose origin coincides with the axis of rotation.

\begin{figure} [ht]
\includegraphics[width=12 cm]{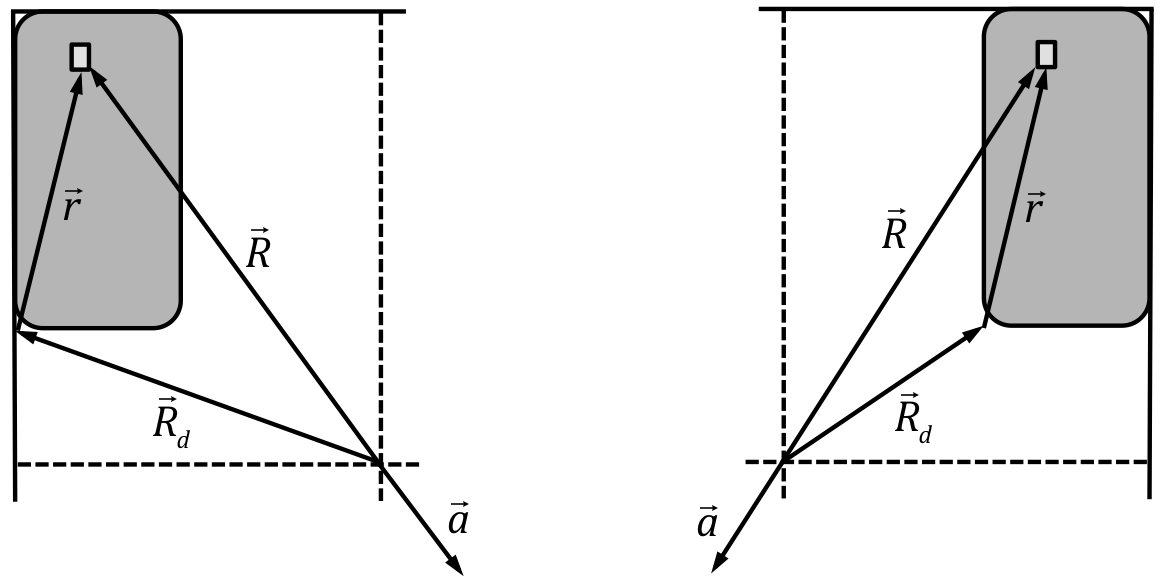}
\centering
\caption{Vector relationships for a host device positioned in the first (right) and second (left) quadrants of a rotating coordinate system.}
\label{fig:VectorRelationships}
\end{figure}

The origin of the host device is placed at location $\vec{R}_ d$  in the global coordinate system and, for simplicity, we constrain its orientation to be one in which the device coordinate axes are aligned with those of the global system. Consequently, the transformation between the two coordinate systems consists only of a single translation by $\vec{R}_ d$, yielding the simplest possible transformation from device coordinates $\vec{r}$ to global coordinates $\vec{R}$, viz. 

\begin{equation}
\label{eq4}
\vec{R} = \vec{r} +\vec{R}_d .
\end{equation}

as illustrated in Fig~\ref{fig:VectorRelationships} for two different device positions $\vec{R}_ d$.

From this the basic procedure for determining $\vec{r}$  is elementary: when the system rotates, the accelerometer reading corresponds to the right-hand side of Eq. \eqref{eq3}. This is solved for $\vec{R}$, at which point application of Eq. \eqref{eq4} yields an estimate for $\vec{r}$.  Algebraic combination of these steps produces 

\begin{equation}
\label{eq5}
\vec{r} = -\frac{1}{\omega^2}\vec{a} -\vec{R}_d .
\end{equation}

\section{Method and Results}

\subsection{Host Devices}

\begin{figure} [ht]
\includegraphics[width=14 cm]{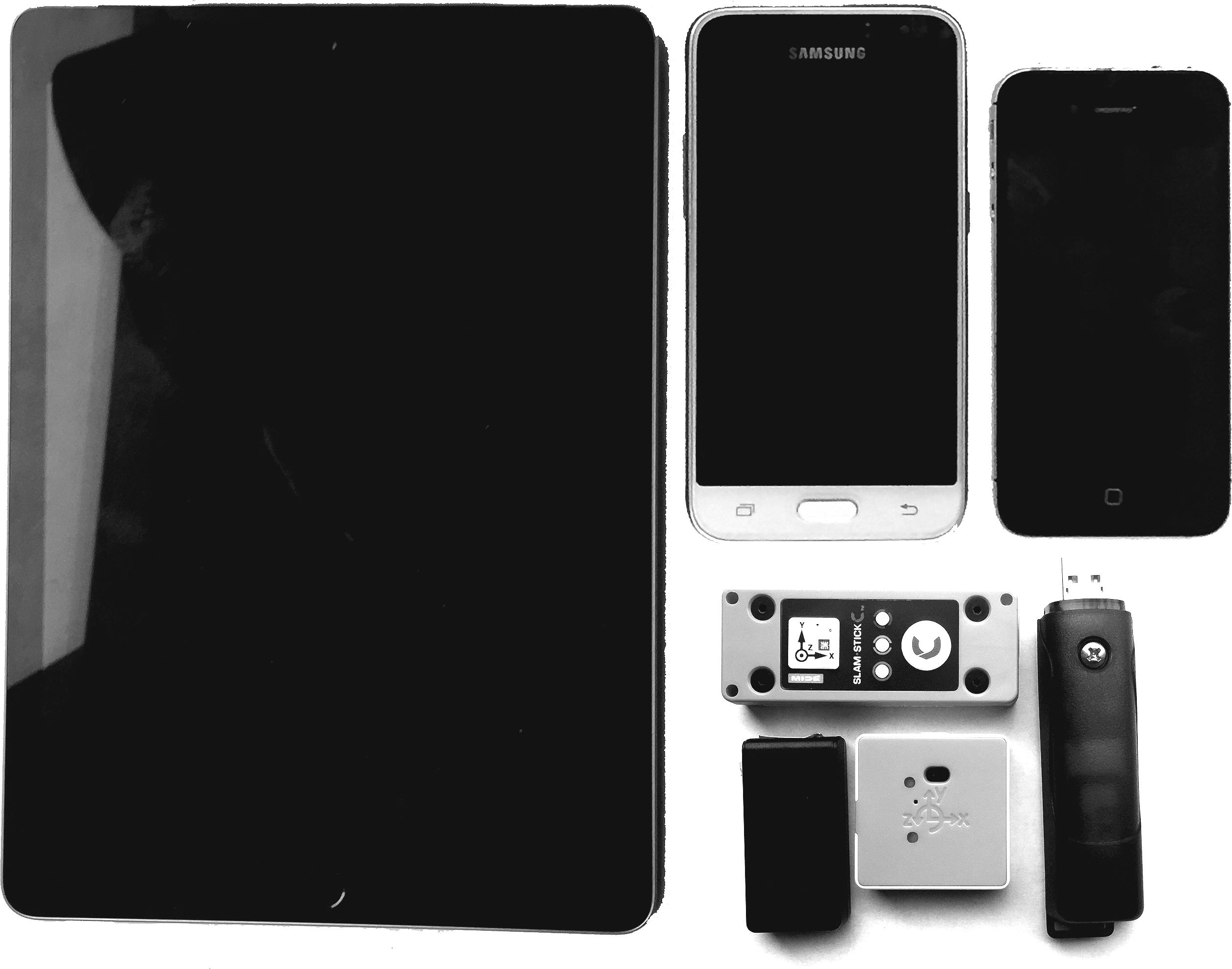}
\centering
\caption{Collection of host devices examined. All are depicted in standard orientation in which the bottom-left corner corresponds to the origin of the local device coordinates. Left-to-right, along the top, we have: iPad 9.7, Samsung Galaxy S5 and Apple iPhone 4S. Left-to-right along the bottom corner we have GCDC x16-mini, PocketLab Voyageur and GCDC x2.2. In the middle is the MIDE SlamStick C. }
\label{fig:HostDevices}
\end{figure}

Investigations were carried out using a variety of host devices whose surfaces areas  range from 13 to 400 cm\textsuperscript{2}, as  tabulated in Table~\ref{tab:HostDevices} and depicted in Fig.~\ref{fig:HostDevices}. These include dedicated accelerometry devices, powered by AA-batteries or by USB-rechargable lithium batteries and controlled via mechanical or magnetic switches, as well as both Android and iOS -based general-purpose smartphones. Although one device streamed data in real-time to a PC, the rest collected and stored data on the device itself which was subsequently transferred to a PC via USB flash-drive or using more recent cloud-based sharing mechanisms. Analysis was performed on the PC using standard spreadsheet applications.

 \begin{table}
 
 \centering
\caption{Host devices used in the experiment. All values rounded to the nearest millimetre. Smartphone dimensions are based on the enclosure without protective cover, and ignoring the protruding buttons. The third dimension perpendicular to the plane of motion is not presented. All data loggers store to a local SD card and provide a USB connection, whereas the multi-sensor streams data wirelessly via the Bluetooth protocol. The X16-mini, the smallest device, is triggered magnetically.}
\label{tab:HostDevices}

 \begin{ruledtabular}

  \begin{tabular} {>{\centering\arraybackslash}m{2.4cm} >{\centering \arraybackslash}m{3.7cm} >{\centering\arraybackslash}m{2.5cm} 
                           >{\centering\arraybackslash}m{3.1cm}}

   \centering host device category & \centering manufacturer & model & dimensions (mm$\times$mm) 
   \\  
  \hline\hline
  
 \multirow{1}{*}{tablets}
& Apple  & iPad 9.7'' & 170 $\times$ 240                 \\ 
 \hline
 
\multirow{3}{*}{\centering smart       phones} 
 & Samsung & SM-J120A & 69 $\times$ 132       	           \\
 & Apple & iPhone 4S       & 59 $\times$ 115         \\ 
\hline

\multirow{3}{2.5cm}{\centering dedicated accelerometry data loggers} 
 & GCDC\footnote{Gulf Coast Data Concepts} & X2-2              & 104 $\times$ 25          \\  
  & MIDE & SlamStick C       &  76 $\times$ 30      \\ 
 & GCDC & X16-mini       &  52 $\times$ 25                \\ 

\hline
 multi-sensors & Pocket Lab & Voyageur    & 38$\times$38  \\ 

 \end{tabular}
\end{ruledtabular}

\end{table}

\begin{figure} [ht]
\includegraphics[width=\textwidth]{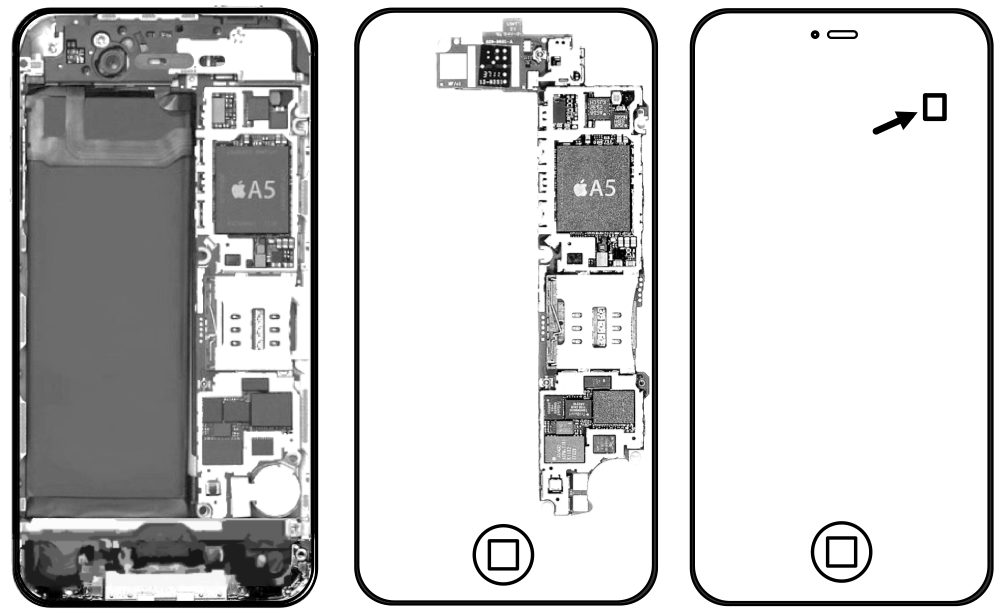}
\caption{Example of a circuit diagram superimposed on a device outline, used for obtaining a direct estimate of accelerometer position. The host device in this case was an iPhone 4S.}
\label{fig:ExampleSchematic}
\end{figure}

Prior to use, each device was subjected to a 6-point tumble calibration procedure\cite{gcdcWEB} \cite{calibration}. Independent values of sensor positions within the device were estimated from manufacturer circuit-board diagrams, an example of which can be found in Fig.~\ref{fig:ExampleSchematic}.

\subsection{Apparatus and general procedure}

The apparatus consists of a 16-inch circular disk made from standard classroom whiteboard material, arranged so as to fit concentrically over the central pin of an inexpensive record player\cite{recordplayer}, as depicted in Fig.~\ref{fig:FourCorners}. A rectangular frame whose inner dimensions are exactly 8.5 inches by 11 inches is attached to the disk, taking special care to ensure that the frame center aligns accurately with that of the disk. Each of the four edges composing the frame are notched at their midpoints. A standard piece of letter-sized paper is close-fitted within the frame, and the notches are used to draw two perpendicular lines across the paper defining the reference axes for the coordinate system, and whose intersection identifies the origin and axis of rotation. 

\begin{figure} [ht]
\includegraphics[width=10 cm]{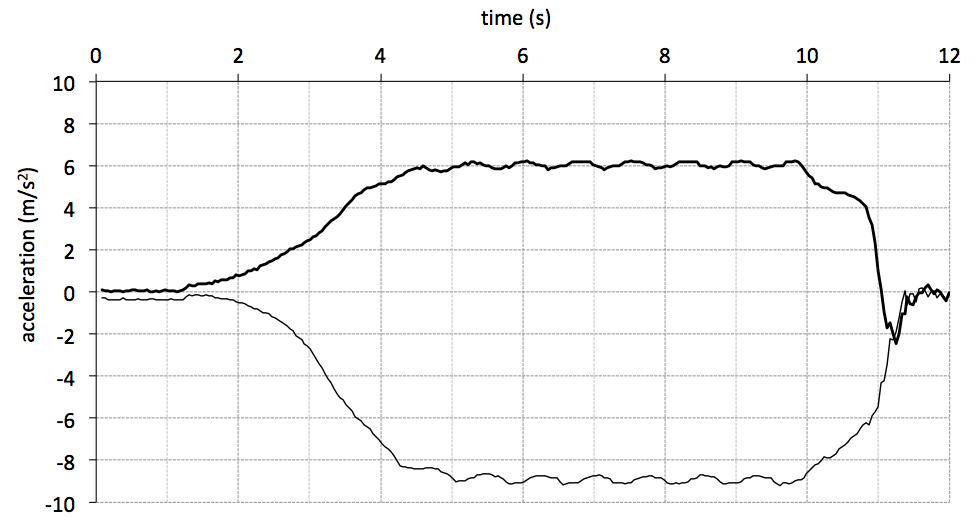}
\centering
\caption{Acceleration components as a function of time for the SlamStick dedicated accelerometry device. The heavier upper line is $a_x$; the lower thinner line is $a_y$ . The device was placed in the second quadrant ( see Fig.~\ref{fig:FourCorners} ), for which a centrally-pointing acceleration vector would necessarily have a positive x-component and a negative y-component. An average is obtained from each component over the time period of uniform circular motion, which in this case was conservatively chosen to be over the interval t = 6 s to t = 9 s. Note the small oscillations around the mean value corresponding to the 78 rpm rotation rate of the turntable. This wobble is due to the weight of the smartphone acting on a turntable bearing not designed for such loads.}
\label{fig:AccelerationGraph}
\end{figure}

Experiments consist of placing the host device at various locations within the frame, maintaining the same orientation with respect to the frame and one device edge in full contact with a frame edge, guaranteeing correct alignment. Data recording is enabled and the turntable is put into motion for a target angular speed of 78 rpm. After 6 seconds, the turntable  is turned off and the recording is terminated. The data is graphed as exemplified in Fig.~\ref{fig:AccelerationGraph}, the constant-acceleration time interval is identified and the average acceleration values along both axes are calculated over this time interval to produce the vector $\vec{a}$.

\subsection{Determination of the intra-device sensor position}

\begin{figure} [ht]
\includegraphics[width=\textwidth]{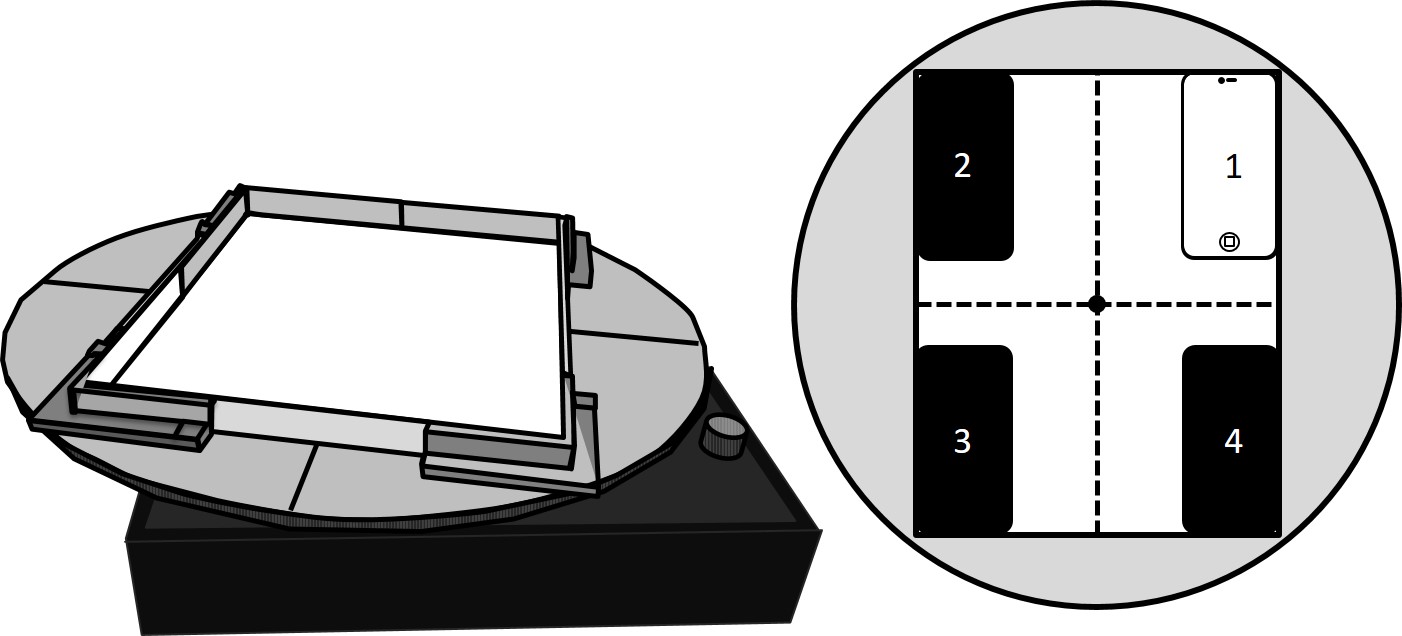}
\centering
\caption{A 16-inch circular disk (left) with a 8.5 inch by 11 inch frame centered upon it. The disk is fitted to a conventional turntable configured to rotate at 78 rpm. The four corner positions of the frame (right) used for obtaining estimates of the intra-device position are indicated.}
\label{fig:FourCorners}
\end{figure}

The above-mentioned Fig.~\ref{fig:FourCorners} also depicts the four corner positions in which the host device is consecutively placed within the frame. In each case the device position $\vec{R}_d$ with respect to the frame coordinates is noted, using the bottom-left-hand-corner convention described above. The system is put into rotation and the general procedure just described is carried out to obtain $\vec{a}$. One then applies Eq. \eqref{eq2} to obtain $\vec{R}$, the position of the sensor with respect to the rotating-frame coordinates, and then, finally, an estimate for $\vec{r}$, the position of the sensor with respect to the device coordinates. Fig.~\ref{fig:VectorRelationships} depicts the relationship between these vectors for the first-quadrant and second-quadrant positions.

\begin{figure} [h]
\includegraphics[width=14 cm]{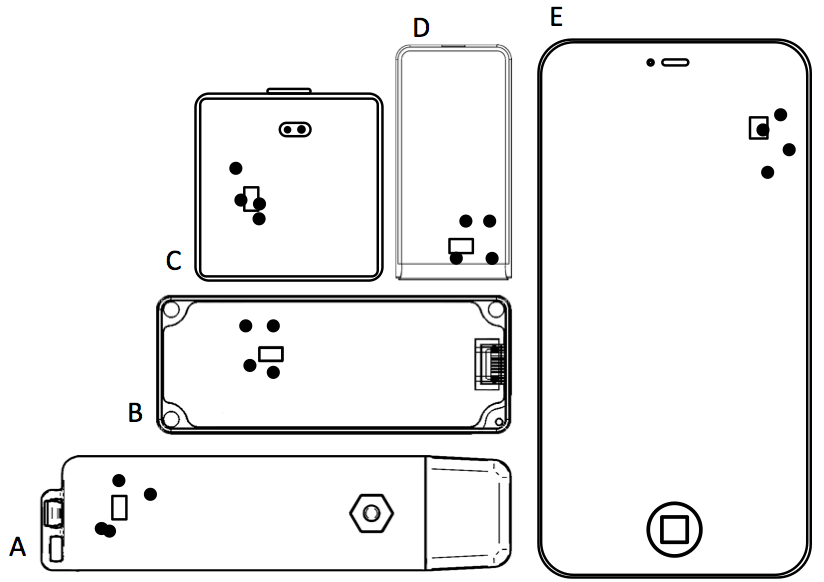}
\centering
\caption{Visualization of experimentally-obtained positions superimposed on host device outlines. The location of the accelerometer obtained by analysis of circuit diagrams is indicated by the small rectangle. Note that the location of the accelerometer within the host device is, in every case, not at the geometrical center. To allow sufficient visual detail, the larger host devices shown in Fig.~\ref{fig:HostDevices} are not shown here. \\
(A.) GCDC X2.2 (B.) MIDE slamstick C (C.) PocketLab Voyager (D.)  GCDC X16-mini (E.) iPhone 4S
}
\label{fig:dots}
\end{figure}

After repeating this procedure for each of the four device positions, an average and standard deviation can be calculated for each component of $\vec{r}$. The results obtained when using the SlamStick device are presented in Table~\ref{tab: SingleDeviceResults}. To improve the statistical confidence, the process can be repeated for as many additional positions as desired, using contact with any portion of the frame edge as a guarantee of correct device alignment. The entire process is repeated for each host device, the results of which are presented in Table~\ref{tab:FinalResults}. Fig.~\ref{fig:dots} depicts the individual vector positions superimposed on the manufacturer's device outline.

 \begin{table}
 
  \centering 
  \caption{Results using MIDE slam stick C device. Device positions $\vec{R}_d$ are chosen according to Fig. \ref{fig:FourCorners}. The averaged $\vec{r}$ position obtained in this case was (24.6,17.3)$\pm$(3.6,6.3) mm.}
\label{tab: SingleDeviceResults}
 \begin{ruledtabular}
 
 \begin{tabular} {>{\centering\arraybackslash}m{1.3cm} >{\centering\arraybackslash}m{2.5cm}  >{\centering\arraybackslash}m{3cm} 
                           >{\centering\arraybackslash}m{2.8cm} >{\centering\arraybackslash}m{2.5cm}}

 quadrant                                                                                                   &      device position $\vec{R}_d$ (mm, mm)     & 
 radial acceleration vector $\vec{a}$  ($\frac{m}{s^2}$, $\frac{m}{s^2}$)  &      sensor position\footnote{frame coordinates} $\vec{R}$ (mm, mm)   &      
 sensor position\footnote{device coordinates} $\vec{r}$ (mm, mm)          \\ 
          
 \hline
 1   &   ( 32, 110 )      &    ( -3.8, -9.0 )   &   ( 57.4, 134.3 )  &  ( 25.5, 24.6 )  \\ 

 2   &   ( -108, 110 )   &   ( 6.0, -9.0 )    &   ( -89.5, 134.8 ) &   ( 18.4, 25.0 )  \\

 3   &   ( -108, 140 )   &   ( 5.9, 8.3 )    &   ( -88.1, -124.5 ) &   ( 19.8, 15.2 )  \\
 
 4   &   ( 32, 140 )      &   ( -3.8, 8.5 )   &   ( 57.1, -126.8 ) &   ( 25.2, 12.9 )  \\ 

\end{tabular}
\end{ruledtabular}

\end{table}

\section{Discussion}

Although Fig.~\ref{fig:dots} is a convincing portrayal of the general validity of our method,
the accuracy of the final positions relative to the standard deviations, as reported in Table~\ref{tab:FinalResults}, underline certain limitations that readers should remain aware of. One limitation relates to the accuracy of the accelerometer sensors themselves. We have tried to minimize such error by applying the linear calibration method mentioned previously, but we remain nevertheless vulnerable both to drift and to non-linearity. We have not undertaken a formal analysis of these error sources nor have reliable estimates for their contribution to the overall error.

A more obvious limitation lies in the experimental conditions themselves, viz. in the use of a low-cost consumer turntable not designed for the degree of loading to which we subjected it ( the accumulated weight of both the frame and the host device ). The result is the wobble clearly seen in the acceleration curves of Fig.~\ref{fig:AccelerationGraph}. The shape is not an exact sinusoid, and it is unclear to what extent averages over multiple cycles of this pattern represent the desired radial acceleration that our theory is based upon. 

A smaller contribution to the error that is nevertheless worth mentioning lies in the estimates of our reference positions which were obtained by superimposing, by eye, various circuit board diagrams over carefully-dimensioned host-device outlines. These estimates, in turn, represent the central position of the sensor whose dimensions are themselves typically on the order of 5 mm.

Although we used the four corners of the frame as our reference device positions, in principle any set of device positions can be used. As long as the device is aligned with a frame edge, then the experimenter can rely on consistent alignment between each measurement. Students may come to realize this on their own, and initiate their own choice of additional device positions to improve the accuracy of their final result and to exercise their ability to think independently.

The use of the four corner device positions remains, however, an excellent starting point for developing the students' ability to map their knowledge of coordinate systems onto the physical representation of the frame itself. With the device placed in the first quadrant, for example, any centrally-pointing vector will have both components negative; in the 3rd quadrant, both components will be positive, and so on. The student can investigate this directly from the raw data, as illustrated in Fig.~\ref{fig:AccelerationGraph} for the 3rd quadrant, or by examining tabulated results, such as in the signs of the third column of Table~\ref{tab: SingleDeviceResults}.

Finally, the experiment also serves as an excellent review and application of vector skills, including the determination of magnitudes, directions and performing additions. Students can reproduce Fig.~\ref{fig:VectorRelationships} to scale, for a given quadrant, or for a device position of their choice, and determine $\vec{r}$ using a graphical technique.

\section{Conclusion}

This present paper has made clear that the position of an accelerometer within its host device, especially in the case of phones, is rarely found at the device's geometric center, but rather dictated by other manufacturing priorities. Understandably, authors of short proof-of-principle papers describing accelerometry experiments may nevertheless consider a treatment of this topic to detract from their primary goal. Indeed, such reservations are the likely reason that a concerted discussion regarding the intra-device sensor position has yet to emerge from recent literature. 

This present paper addresses this situation firstly by pointing out that, even in ignorance of the actual intra-device sensor position, there is nevertheless a simple method for quickly estimating the size of the associated experimental error. Secondly, should the size of the error warrant it, a procedure for reliably determining the sensor position has been proposed and demonstrated across a wide variety of host devices. Finally, the procedure itself, far from being an additional burden on an experimental program, could be developed into a valuable pedagogical tool with the potential to generate excitement and anticipation as the repeated application of the classical principles of radial acceleration reveal, with increasing confidence, a hidden property behind the digital devices that most students and practitioners interact with on a daily basis.

\section{Acknowledgements}

The first author would like to thank to Peter Hinrichsen for encouragement and for sharing an initial collection of relevant research literature. This work was funded in part by Qu\'{e}bec's \textit{Minist\`{e}re de l'\'{E}ducation et de l'Enseignement sup\'{e}rieure}, through contributions from the Canada-Québec Agreement on Minority-Language Education and Second-Language Instruction.

\begin{table}
  \centering 
\caption{Comparison of experimentally obtained estimates of intra-device position with those estimated from manufacturer's schematics. All geometrical values save deviation rounded to the nearest millimeter. The experimental position is the average of 4 estimates. The standard deviation is the corrected sample standard deviation, while the standard score is the ratio of deviation to this standard deviation. }

\label{tab:FinalResults}
 \begin{ruledtabular}
 
 \begin{tabular} {>{\centering\arraybackslash}m{2.2cm} >{\centering\arraybackslash}m{2.5cm}  >{\centering\arraybackslash}m{3cm} 
                           >{\centering\arraybackslash}m{2.8cm} >{\centering\arraybackslash}m{2.5cm}>{\centering\arraybackslash}m{2cm} }

device & position from schematic (mm, mm)  & experimental position (mm, mm)   &     deviation  (mm, mm) &  standard deviation $s$ (mm, mm)   &      standard score $z$   \\

 \hline
 iPad   &   ( 150, 156 )      &    ( 150, 160 )   &   ( 0.5, -4 )& ( 4, 2 )    & ( 0.1, -1.8 )  \\ 
 
 iPhone 4S  &   ( 48, 97 )   &   ( 51, 93 )    &  ( -3 , 4 ) & ( 3, 5 ) &    ( -1.0, 1.2 )  \\

 X2-2   &   ( 84, 14 )      &   ( 83, 13 )   &   ( 0.8, 0.5 )  & ( 5, 6 )   & ( 0.2, 0.1 ) \\ 
 
 X16-mini   &   ( 43, 14 )      &   ( 42, 18 )   &  ( 1.2, -4 )  & ( 5, 4 ) &    ( 0.2, -0.9 ) \\ 

Galaxy S5 &   ( 58, 93)      &    ( 60, 101 )   &   ( -3, -8 ) & ( 5, 3 )    & ( -1.3, -2 )  \\
 
 Slam Stick C   &   ( 25, 17 )      &   ( 22, 19 )   &   ( 3, -1.6 )  &( 4, 6 ) &    ( 0.8, -0.3 ) \\ 
 
 Voyageur   &   ( 13, 19 )      &   ( 12, 20 )   &    ( 1.5, -1.5 )  & ( 3, 4 ) &  ( 0.4, -0.4  ) \\ 

\end{tabular}
\end{ruledtabular}

\end{table}


\begin{thebibliography}{9}
\bibitem{footnote_rot} 
Although some points may share the same magnitude or the same direction, no two points are identical in both characteristics.
\bibitem{footnote_gyro} 
In practise, gyroscopes are used to detect changes in direction, and accelerometers only for changes in linear acceleration, but the point still holds, given that accelerometer readings still need to be corrected for such changes in direction.
\bibitem{footnote_klein} 
The uncertainty due to true sensor position would lie along the short side of the device, given the experimental conditions described.  In the absence of details regarding exactly how radial position was measured, we estimate the uncertainty in position to be the half-width of the short side, which for the iPod Touch is roughly 30 mm. \par To be fair, the estimation of damping model parameters, which was the goal of the paper in question, is not affected by the value of the radial distance, as the latter remains a constant with no variation.
\end{thebibliography}
\end{document}